\documentclass[a4paper,10pt,aps,amsfonts,amsmath,nofootinbib,showpacs,byrevtex,showkeys%
,notitlepage%
]{revtex4-1}

\usepackage{hyperref,graphicx,bm}

\renewcommand{\vec}[1]{\ensuremath{\bm{\mathrm{#1}}}}%
\newcommand*{\uvec}[1]{\ensuremath{\Hat{\bm{\mathrm{#1}}}}}%
\newcommand{\I}{\ensuremath{\mathrm{i}}}%

\newcommand*{\dbar}[1][]{\mathrm{d}\mkern-7mu\mathchar'26\mkern-1mu^{#1}}

\newcommand*{\be}{\begin{equation}}
\newcommand*{\ee}{\end{equation}}

\newcommand*{\vev}[1]{\left< #1 \right>}
\newcommand*{\sigmacoul}{\sigma_{\textsc{c}}}

\DeclareMathOperator{\Det}{Det}

\newcommand*{\vx}{\vec{x}}
\newcommand*{\vy}{\vec{y}}
\newcommand*{\vp}{\vec{p}}
\newcommand*{\vq}{\vec{q}}
\newcommand*{\vk}{\vec{k}}
\newcommand*{\vl}{\vec{\ell}}
\newcommand*{\uvp}{\uvec{p}}

\newcommand*{\uvk}{\uvec{k}}

\newcommand{\Nc}{{N_{\mathrm{c}}}}

\providecommand*{\eqcolon}{=\mathrel{\mathop:}}

\renewcommand*{\d}[1][]{\mathrm{d}^{#1}} 
\newcommand*{\abs}[1]{\ensuremath{\lvert#1\rvert}}

\newcommand*{\calD}{\ensuremath{\mathcal{D}}}
\newcommand*{\calJ}{\ensuremath{\mathcal{J}}}

\begin{document}

\title{The ghost-gluon vertex in Hamiltonian Yang--Mills theory in Coulomb gauge}
\author{Davide R.~Campagnari}
\author{Hugo Reinhardt}
\affiliation{Institut f\"ur Theoretische Physik, Universit\"at T\"ubingen, Auf der Morgenstelle 14, 72076 T\"ubingen, Germany}
\date{23 November 2011}

\begin{abstract}
The Dyson--Schwinger equation for the ghost-gluon vertex of the Hamiltonian approach to
Yang--Mills theory in Coulomb gauge is solved at one-loop level using as input the
non-perturbative ghost and gluon propagators previously determined within the variational
approach. The obtained ghost-gluon vertex is IR finite but IR enhanced compared to the
bare one by 15\% to 25\%, depending on the kinematical momentum regime.
\end{abstract}


\maketitle

\section{Introduction}

In recent years there have been intensive studies of continuum Yang--Mills theory in the
non-perturbative regime. Most of these investigations were carried out either in Landau
gauge, using Dyson--Schwinger equations (DSEs) (for reviews see
Refs.~\citep{AlkSme01,Fis06,Sme08,BinPap09}) or functional renormalization group (FRG)
flow equations (for reviews see Refs.~\citep{Paw05,Gie06}), or in Coulomb gauge, using
DSEs \citep{WatRei}. In addition, variational
\citep{Sch85,SzcSwa01,FeuRei04,ReiFeu05,SchLedRei06,EppReiSch07,Rei08,CamRei10}
or FRG methods \citep{Led+10} were used in the Hamiltonian formulation of Yang--Mills
theory in Coulomb gauge. All those various approaches have in common that the infrared
(IR) sector of the theory is dominated by the ghost degrees of freedom, which has been
referred to as ``ghost dominance''. For this reason the ghost-gluon vertex is of crucial
importance in these approaches. Inspired by Taylor's work \citep{Tay71} showing the
non-renormalization of the ghost-gluon vertex in Landau gauge, in all the approaches
mentioned above it was tacitly assumed that the ghost-gluon vertex is bare. In
Ref.~\citep{Sch+05} a semi-perturbative calculation of the ghost-gluon vertex was
carried out and it was found that its dressing is small even in the IR, a result which is also
found on the lattice \citep{CucMenMih04,CucMaaMen08}.
Although Taylor's proof of non-renormalization formally applies also to the ghost-gluon
vertex in the Hamiltonian formulation of Yang--Mils theory in Coulomb gauge, the results
for the dressing of the ghost-gluon vertex obtained in the functional integral formulation
of Yang--Mills theory in Landau gauge \citep{Sch+05} cannot be assumed to remain valid
also in the Hamiltonian approach in Coulomb gauge. Furthermore, recently it was found
within the FRG-approach in Landau gauge that the dressing of the ghost-gluon vertex
becomes crucial at high temperatures \citep{FisPawX}. Since the high-temperature limit
of the 4-dimensional Yang--Mills theory is essentially the 3-dimensional Euclidean Yang--Mills
theory and the latter provides an approximation to the Yang--Mills vacuum wave functional in
$3+1$ dimensions\footnote{%
More precisely, the functional integral of 3-dimensional Euclidean Yang--Mills theory in
Landau gauge can be interpreted as the functional integral of the vacuum expectation
value of the Hamiltonian approach in Coulomb gauge in $3+1$ dimensions with a vacuum
wave functional given by $\psi[A]\sim\exp(-\frac12 S_{\textsc{ym}}[A])$, where
$S_{\textsc{ym}}[A]$ is the classical action of 3-dimensional Euclidean Yang--Mills
theory. This wave functional was shown to provide a decent approximation to the true
Yang--Mills vacuum wave functional in the mid-momentum regime \cite{Quandt:2010yq}.}%end footnote
 \citep{Gre79}, the dressing of the ghost-gluon vertex of the
Hamiltonian approach in Coulomb gauge should be expected to be also substantial.
Therefore, in the present paper we investigate the ghost-gluon vertex of the Hamiltonian
approach to Yang--Mills theory in Coulomb gauge. We will solve the DSE for the ghost-gluon
vertex using as input the non-perturbative ghost and gluon propagators previously obtained
in the variational approach.

In Sec.~\ref{sec:props} we briefly summarize the basic ingredients of the Hamiltonian
approach to Yang--Mills theory in Coulomb gauge and also present the results obtained
for the ghost and gluon propagators. In Sec.~\ref{sec:ggvdse} we give a short derivation
of the DSE for the ghost-gluon vertex, and introduce the truncation scheme. Our numerical
results are presented in Sec.~\ref{sec:res}. Some concluding remarks are given in
Sec.~\ref{sec:conc}.

%%%%%%%%%%%%%%%%%%%%%%%%%%%%%%%%%%%%%%%%%%%%%%%%%%%%%%%%%%%%%%%%%%%%%%%%%%%%%%%%%%%%%%%%%
%%%%%%%%%%%%%%%%%%%%%%%%%%%%%%%%%%%%%%%%%%%%%%%%%%%%%%%%%%%%%%%%%%%%%%%%%%%%%%%%%%%%%%%%%

\section{\label{sec:props}Equal-time propagators in Coulomb gauge}

The Hamiltonian approach to Yang--Mills theory in Coulomb gauge is based on canonical
quantization in Weyl gauge $A_0^a=0$, and results in a Schr\"odinger equation, which has
to be solved for the vacuum wave functional $\psi[A]=\langle A \vert 0\rangle$ of the
transverse gauge field $\partial_i A_i^a=0$. Once $\psi[A]$ is known all static
(time-independent) Green's functions can be evaluated. In Refs.~\citep{FeuRei04,EppReiSch07}
the Yang--Mills Schr\"odinger equation was solved in an approximate fashion using the
variational principle and assuming Gaussian-type trial wave functionals.

In the Hamiltonian formulation of Yang--Mills theory, Coulomb gauge can be implemented
in the expectation value of any functional $K[A]$ of the (spatial components of the) gauge
field $A$ by the Faddeev--Popov method, which results in
\be\label{vev1}
\vev{K[A]} = \int_\Omega \calD A \: \calJ_A \: \abs{\psi[A]}^2 \: K[A] \, .
\ee
Here, $\calJ_A = \Det (G^{-1}_A)$ is the Faddeev--Popov determinant with
\be\label{ggvdse0}
G_A^{-1}{}^{ab}(\vx,\vy) = \bigl( - \delta^{ab} \partial^2 - g f^{acb} A_i^c(\vx) \partial_i \bigr) \delta(\vx-\vy)
\ee
being the Faddeev--Popov operator. Furthermore, $g$ is the coupling constant and $f^{acb}$
are the structure constants of the $\mathfrak{su}(\Nc)$ algebra. The functional integration
in Eq.~(\ref{vev1}) extends over transverse field configurations restricted to the
first Gribov region $\Omega$ or, more precisely, to the fundamental modular region
\cite{Zwa94}. Moreover, we assume that the wave functional $\psi[A]$ is properly
normalized, such that $\vev{1} = 1$.

To simplify the bookkeeping we will use the compact notation $A^{a_1}_{i_1}(\vx_1)\equiv A(1)$
and assume that a repeated label means summation over colour and spatial indices along with
integration over the spatial coordinates,
\be\label{notation}
A \cdot B = A(1) B(1) = \int \d[3]x \: A_i^a(\vx) B_i^a(\vx) .
\ee
We use the same convention for indices referring to the ghost field except that the label ``1''
represents only the colour index $a_1$ and the spatial coordinate $\vx_1$.

The gluon propagator $D$ and the ghost propagator $G$ are defined by the expectation values
\be\label{ggvdse0c}
D(1,2) = \vev{ A(1) A(2) }, \qquad G(1,2) = \vev{ G_A(1,2) } .
\ee
In momentum space we express the gluon propagator by the gluon energy $\Omega(\vp)$
\be\label{ggvdse4b}
\vev{ A_i^a(\vp) A_j^b(\vq) } \eqcolon \delta^{ab} (2\pi)^3 \delta(\vp+\vq) \frac{t_{ij}(\vp)}{2\Omega(\vp)} ,
\ee
where $t_{ij}(\vp) = \delta_{ij}-p_i p_j/\vp^2$ is the transverse projector in momentum
space. Furthermore the ghost propagator can be represented as
\be\label{ghostprop}
G^{ab}(\vp) = \delta^{ab} \frac{d(\vp)}{g \vp^2} ,
\ee
where $d(\vp)$ is the ghost form factor. Assuming the so-called horizon condition
$d^{-1}(0)=0$, the results obtained with Gaussian-type wave functionals \cite{EppReiSch07}
show an IR diverging gluon energy $\Omega(\vp)$ which can be fitted with Gribov's
formula \citep{Gri78}
\be\label{numres1}
\Omega(\vp^2) = \sqrt{\vp^2+\frac{m^4_A}{\vp^2}} \, ,
\ee
with a mass parameter $m_A^2\simeq 0.6 \, \sigmacoul$ (for $\Nc=2$), where
$\sigmacoul$ is the Coulomb string tension, i.e.\ the coefficient of the linear
term of the non-Abelian Coulomb potential. These results compare favourably
with recent lattice calculations \cite{BurQuaRei09}; in particular, the infrared regime
of the gluon energy is correctly reproduced.

%%%%%%%%%%%%%%%%%%%%%%%%%%%%%%%%%%%%%%%%%%%%%%%%%%%%%%%%%%%%%%%%%%%%%%%%%%%%%%%%%%%%%%%%%
%%%%%%%%%%%%%%%%%%%%%%%%%%%%%%%%%%%%%%%%%%%%%%%%%%%%%%%%%%%%%%%%%%%%%%%%%%%%%%%%%%%%%%%%%

\section{\label{sec:ggvdse}Derivation of the ghost-gluon vertex DSE}

In the functional integral formulation of Yang--Mills theory in Landau gauge
the DSE for the ghost-gluon vertex has been known for quite some time, see e.g.\ Ref.~\citep{Smi74}.
In the Hamiltonian approach in Coulomb gauge the DSE for the ghost-gluon vertex was derived in 
Ref.~\citep{CamRei10}. We briefly summarize this derivation in order to fix our notation.

\begin{figure}
\centering\includegraphics[width=.6\linewidth]{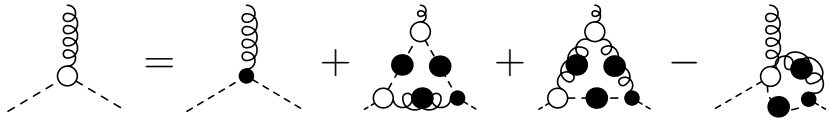}
\caption{\label{fig:ggvDSE}Diagrammatic representation of the DSE \eqref{ggvdse3}. Small
filled dots represent connected Green's functions, and small empty dots proper vertex functions.}
\end{figure}

The Faddeev--Popov operator Eq.~\eqref{ggvdse0}
can be inverted to give the following operator identity for the ghost operator $G_A$
\be\label{ggvdse1}
G_A(1,2) = G_0(1,2) - G_A(1,4) A(5) \, \widetilde{\Gamma}_0(5;4,6) \, G_0(6,2).
\ee
Here the bare ghost-gluon vertex $\widetilde{\Gamma}_0$ is defined by
\be\label{ggvdse0a}
\widetilde{\Gamma}_0(1;2,3) = \frac{\delta G_A^{-1}(2,3)}{\delta A(1)} \, ,
\ee
and agrees with the lowest order term in the perturbative expansion of the full vertex
$\widetilde{\Gamma}$ defined by
\be\label{ggvdse0b}
\vev{A(3) G_A(1,2) } \eqcolon - D(3,3') \, G(1,1') \, G(2',2) \, \widetilde{\Gamma}(3';1',2') .
\ee
Taking the v.e.v.\ of Eq.~\eqref{ggvdse1} and using Eq.~\eqref{ggvdse0b} one obtains
the DSE for the ghost propagator $G$, which reads
\be\label{ghostdse}
G^{-1}(1,2) = G_0^{-1}(1,2) - \widetilde{\Gamma}(3;1,4) D(3,3') G(4,4') \widetilde{\Gamma}_0(3';4',2) .
\ee
If Eq.~\eqref{ggvdse1} is
multiplied by a gauge field $A(3)$ before taking the expectation value, we are led to
\be\label{ggvdse2}
\vev{ G_A(1,2) A(3) } = - \widetilde{\Gamma}_0(5;4,6) \, G_0(6,2)  \vev{G_A(1,4) A(5) A(3)} ,
\ee
where we have used $\vev{A}=0$. For the left-hand side of Eq.~\eqref{ggvdse2} the
relation \eqref{ggvdse0b} can be used, while the expectation value on the right-hand
side can be expanded in terms of vertex functions and propagators \citep{CamRei10}. By
using the DSE \eqref{ghostdse} for the ghost propagator, the DSE \eqref{ggvdse2} for
the ghost-gluon vertex becomes
\be\label{ggvdse3}
\begin{split}
\widetilde{\Gamma}(1;2,3) = \widetilde{\Gamma}_0(1;2,3)
&+ \widetilde{\Gamma}(1;4,5) G(4',4) G(5,5') \widetilde{\Gamma}(6';5',3) D(6,6') \widetilde{\Gamma}_0(6;2,4') \\
&+ \Gamma(1,4,5) D(4,4') D(5,5') \widetilde{\Gamma}(4';2,6) G(6,6') \widetilde{\Gamma}_0(5';6',3) \\
&- \widetilde{\Gamma}(1,5;2,4) D(5,5') G(4,4') \widetilde{\Gamma}_0(5';4',3) ,
\end{split}
\ee
where $\Gamma(1,4,5)$ is the three-gluon vertex, and $\widetilde{\Gamma}(1,5;2,4)$ is
the ghost-gluon scattering kernel. Equation \eqref{ggvdse3} is represented diagrammatically
in Fig.~\ref{fig:ggvDSE}.
Note that the vacuum wave functional does not explicitly enter this equation,
but only implicitly via the various propagators, in particular the gluon propagator.

Equation~\eqref{ggvdse3} must be truncated to be feasible; in a first step we will discard the
proper four-point function. Even then, a fully self-consistent solution of the resulting equation
together with the DSEs for the propagators is still very expensive and beyond the scope of
this work. To get a first estimate of the size of the dressing of the ghost-gluon vertex we will keep the full,
non-perturbative propagators but approximate the vertices in the loop terms by their
bare form. If the usual assumption of a bare ghost-gluon vertex is justified,
the corrections to the bare vertex from the one-loop terms should turn out to be small.

The bare three-gluon vertex is given by the three-gluon kernel $\gamma_3$ in the exponent
of the vacuum wave functional \cite{CamRei10}
\be\label{vacansatz}
\psi[A] = \exp\left\{ - \frac12 \omega A^2 - \frac{1}{3! 2} \gamma_3 A^3 - \dots \right\}
\ee
which is found from the variational calculation to be given by \cite{CamRei10}
\be\label{3gv}
\gamma^{abc}_{ijk}(\vp,\vq,\vk) = \frac{2 f^{abc} T_{ijk}(\vp,\vq,\vk) }{\Omega(\vp)+\Omega(\vq)+\Omega(\vk)} ,
\ee
where $T$ is the Lorentz structure of the three-gluon coupling in the Hamiltonian
\be\label{lorentz3gv}
T_{ijk}(\vp,\vq,\vk) = \I g \bigl[ \delta_{ij} (p-q)_k + \delta_{jk} (q-k)_i + \delta_{ki} (k-p)_j \bigr] .
\ee
Note that $\gamma_3$ Eq.~\eqref{3gv} is the bare part of the three-gluon vertex $\Gamma_3$
in the sense that it is the leading term in the DSE for $\Gamma_3$. However, $\gamma_3$ Eq.~\eqref{3gv}
is not the lowest-order perturbative vertex, which is obtained from Eq.~\eqref{3gv} when
the gluon energy $\Omega(\vp)$ is replaced by its perturbative counterpart $\abs{\vp}$.

After implementing the truncations explained above and extracting the colour structure,
Eq.~\eqref{ggvdse3} reads in momentum space
\be\label{ggvdse4}
\begin{split}
\widetilde{\Gamma}_i(\vk;\vp,\vq) ={}& \I g \, t_{ij}(\vk) p_j
- t_{ij}(\vk) \frac{\Nc}{2} \int \dbar\ell \:
\bigl(\I g p_m\bigr) \, G(\vl) \, \bigl(\I g \ell_j\bigr) \, G(\vl+\vk) \, \bigl(\I g (\ell+k)_n \bigr)\:
\frac{t_{mn}(\vl-\vp)}{2\Omega(\vl-\vp)} \\
&-
t_{ij}(\vk) \frac{\Nc}{2} \int \dbar\ell \:
\bigl(\I g p_l\bigr) \: \frac{t_{lm}(\vl)}{2\Omega(\vl)} \:
\frac{2 T_{jmn}(\vk,\vl,-\vl-\vk)}{\Omega(\vk)+\Omega(\vl)+\Omega(\vk+\vl)}
\frac{t_{nk}(\vl+\vk)}{2\Omega(\vl+\vk)} \: \bigl(\I g (p-\ell)_k\bigr) \, G(\vl-\vp) ,
\end{split}
\ee
where we have introduced the notation $\dbar\ell \equiv \d[3]\ell/(2\pi)^3$. In
Eq.~\eqref{ggvdse4} $\vk$, $\vp$, and $\vq$ are respectively the momenta of the gluon
and the incoming and outgoing ghost, and momentum conservation $\vk+\vp+\vq=0$ is
implicitly understood.

We parameterize the full ghost-gluon vertex by a dressing function $h$ as
\be\label{ggvff1}
\widetilde{\Gamma}_i(\vk;\vp,\vq) = \I g \, t_{ij}(\vk) p_j \bigl[ 1+ h(k^2;p^2,q^2) \bigr] .
\ee
Let us stress that in the present Hamiltonian approach the Coulomb gauge condition is
exactly implemented, and the functional integral of the scalar product of the Hilbert
space is strictly restricted to (spatially) transverse gauge fields. Therefore, in the
present case the ghost-gluon vertex cannot develop a longitudinal part.

Contracting Eqs.~\eqref{ggvdse4} and \eqref{ggvff1} with $p_i$ and dividing both sides
by $p_i p_j t_{ij}(\vk)$ we obtain\footnote{The limit $\uvp\cdot\uvk \to \pm1$ is finite.}
\be\label{ggvdse6}
h(k^2;p^2,q^2) = I_1(k^2;p^2,q^2) + g I_2(k^2;p^2,q^2),
\ee
where
\begin{subequations}\label{ggvdse5}
\be\label{ggvdse5a}
I_1 = \frac{\Nc}{4 p^2 [1-(\uvk\cdot\uvp)^2]}
\int \dbar\ell \: p_i t_{ij}(\vk)\ell_j \, \frac{d(\vl)}{\vl^2} \frac{d(\vl+\vk)}{(\vl+\vk)^2} \frac{p_m (p+k)_n \, t_{mn}(\vl-\vp)}{\Omega(\vl-\vp)}
\ee
is the contribution of the diagram with three ghost-gluon vertices, and 
\be\label{ggvdse5b}
I_2 = \frac{\Nc}{4 p^2 [1-(\uvk\cdot\uvp)^2]} \int \dbar\ell \:
\frac{p_i p_j (p+k)_k t_{il}(\vk) t_{jm}(\vl) t_{kn}(\vl+\vk) T_{lmn}(\vk,\vl,-\vl-\vk)}%
      {\Omega(\vl)\Omega(\vl+\vk) [\Omega(\vk)+\Omega(\vl)+\Omega(\vl+\vk)]} \: \frac{d(\vl-\vp)}{(\vl-\vp)^2}
\ee
\end{subequations}
is the contribution of the diagram containing the three-gluon vertex; the latter is
multiplied by the coupling constant $g$ due to the parameterization of the ghost propagator,
see Eq.~\eqref{ghostprop}.
It should be remarked here that the integrals Eqs.~\eqref{ggvdse5} are UV finite and need
not to be renormalized.
In the subsequent numerical calculations we will
consider two classes of kinematic configurations, namely
\be\label{kinconf}
h(p^2;p^2,x p^2) \qquad \text{and} \qquad h(x p^2;p^2,p^2),
\ee
where $x$ is restricted to the interval $x\in[0,4]$ due to momentum conservation.

Before concluding this section we investigate the IR limit of the ghost-gluon vertex.
In the IR the ghost and gluon propagator can be parameterized as
\be\label{irprops}
d(\vp\to0) \sim \frac{m_c^\beta}{\abs{\vp}^\beta} , \qquad
\Omega(\vp\to0) \sim \frac{m_A^{1+\alpha}}{\abs{\vp}^\alpha} ,
\ee
with $\beta>0$ to fulfil the horizon condition $d^{-1}(0)=0$.
These IR exponents obey the sum rule
$1+\alpha=2\beta$ \citep{SchLedRei06,Zwa02,LerSme02}.
Inserting the IR ansatzes Eq.~\eqref{irprops} into Eqs.~\eqref{ggvdse5} we find that
in the limit of vanishing momenta the form factor of the ghost gluon vertex approaches
the finite expression
\be\label{irvalue}
h(0;0,0) = \frac{\Nc}{24(1+\beta) \pi^2} \left(\frac{m_c}{m_A}\right)^{2\beta} .
\ee
A fit to the numerical data \cite {EppReiSch07} for the propagators with $\Nc=2$ and
the solution $\beta=1$ yields $m_c \simeq 4.97 \sqrt{\smash[b]{\sigmacoul}}$ while
$m_A$ was given below Eq.~\eqref{numres1}. Plugging these values into Eq.~\eqref{irvalue}
yields an IR value of 0.174 for the ghost-gluon-vertex form factor $h$. The numerical data
shown in the next section confirm this value.

%%%%%%%%%%%%%%%%%%%%%%%%%%%%%%%%%%%%%%%%%%%%%%%%%%%%%%%%%%%%%%%%%%%%%%%%%%%%%%%%%%%%%%%%%
%%%%%%%%%%%%%%%%%%%%%%%%%%%%%%%%%%%%%%%%%%%%%%%%%%%%%%%%%%%%%%%%%%%%%%%%%%%%%%%%%%%%%%%%%

\section{\label{sec:res}Numerical results}

For the ghost and gluon propagators we use the results obtained previously in the
variational approach \citep{EppReiSch07} as input.
In particular, for the gluon propagator we use the Gribov formula [Eq.~\eqref{numres1}],
while for the ghost form factor [Eq.~\eqref{ghostprop}] we use the parameterization \citep{CamRei10,WatRei10}
\be\label{numres2}
d(\vp) = a \sqrt{\frac{1}{\vp^2/\sigmacoul}+\frac{1}{\ln(\vp^2/\sigmacoul+c)}} \, ,
\ee
with $a=4.97$ and $c=16$. The factor $g$ has been absorbed in the ghost propagator as
explained in Refs.~\citep{FeuRei04,EppReiSch07}. This has the advantage that the coupling
constant $g$ disappears from the coupled system of equations, as long as gluonic vertices
are ignored. However, since the diagram containing the three-gluon vertex has a prefactor
$g^2$ but only a single ghost propagator in the integral, in Eq.~\eqref{ggvdse6} there
remains an explicit factor $g$. We will take the coupling at the
renormalization point $\mu=2.4\sqrt{\smash[b]{\sigmacoul}}$ \citep{EppReiSch07} $g_r=3.5$.

\begin{figure}
\centering
\includegraphics[width=.45\linewidth]{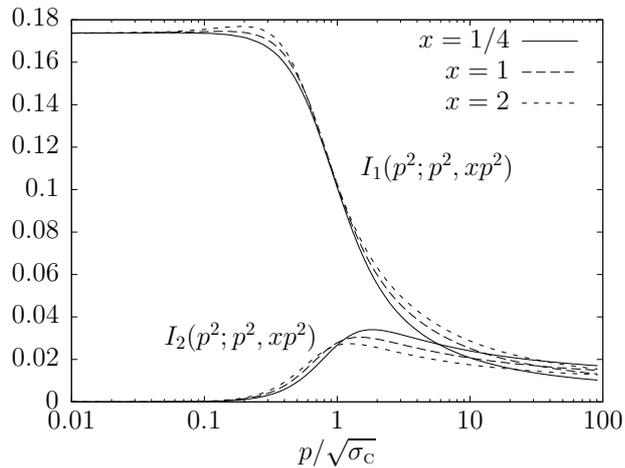}
\caption{\label{fig:kp_sep}Numerical results for the two integrals contributing to the
form factor of the ghost-gluon vertex $h(p^2;p^2,xp^2)$, see Eqs.~\eqref{ggvdse6} and
\eqref{ggvdse5}.}
\end{figure}

\begin{figure}
\centering
\includegraphics[width=.45\linewidth]{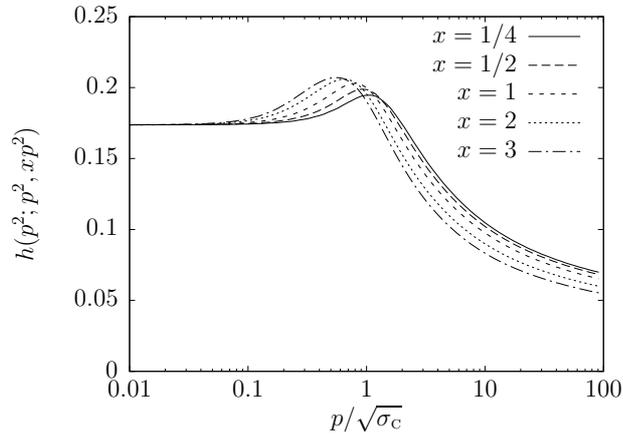}
\caption{\label{fig:kp}Numerical results for the form factor $h(p^2;p^2,xp^2)$ of the
ghost-gluon vertex with equal ghost and gluon momentum.}
\end{figure}

Figure \ref{fig:kp_sep} shows the values of the two integrals $I_1$ and $I_2$ separately,
for equal ghost and gluon momentum.
The integral $I_1$ involving three ghost-gluon vertices approaches the IR value evaluated
in Eq.~\eqref{irvalue}, showing a modest dependence on the kinematic configuration.
At high momenta it drops off logarithmically, due to the anomalous dimension of the ghost
propagator. The integral $I_2$ involving the three-gluon vertex vanishes in the deep IR,
due to the IR divergence of the gluon energy $\Omega(\vk)$ in the denominator, see Eqs.~\eqref{ggvdse4} and
\eqref{ggvdse5}, and drops off in the UV more slowly than the ghost-loop term
($1/\surd\ln p$ instead of $1/\ln p$), since it contains only one ghost propagator.
Figure \eqref{fig:kp} shows the total form factor $h(p^2;p^2,xp^2)$.

\begin{figure}
\centering
\includegraphics[width=.45\linewidth]{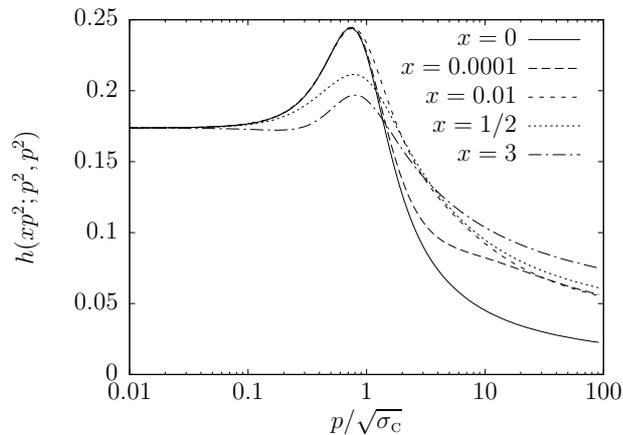}
\caption{\label{fig:pq}Numerical results for the form factor $h(xp^2;p^2,p^2)$
with equal incoming and outgoing ghost. Explanation see text.}
\end{figure}

More interesting is the kinematic configuration where the ghost legs have momenta
of equal magnitude and the gluon momentum is varied, see Fig.~\ref{fig:pq}.
One observes that the maximum in the mid-momentum regime gets stronger as the gluon
momentum approaches zero. If the gluon momentum is very small but not zero (dashed line
with $x=10^{-4}$ in Fig.~\ref{fig:pq}), the form factor is almost indistinguishable
from the case of vanishing gluon momentum in the IR and in the mid-momentum regime;
however, at higher momenta the term involving the three-gluon vertex dominates, and we
observe again the $(\ln p)^{-1/2}$ behaviour.

\begin{figure}
\centering
\includegraphics[width=.45\linewidth]{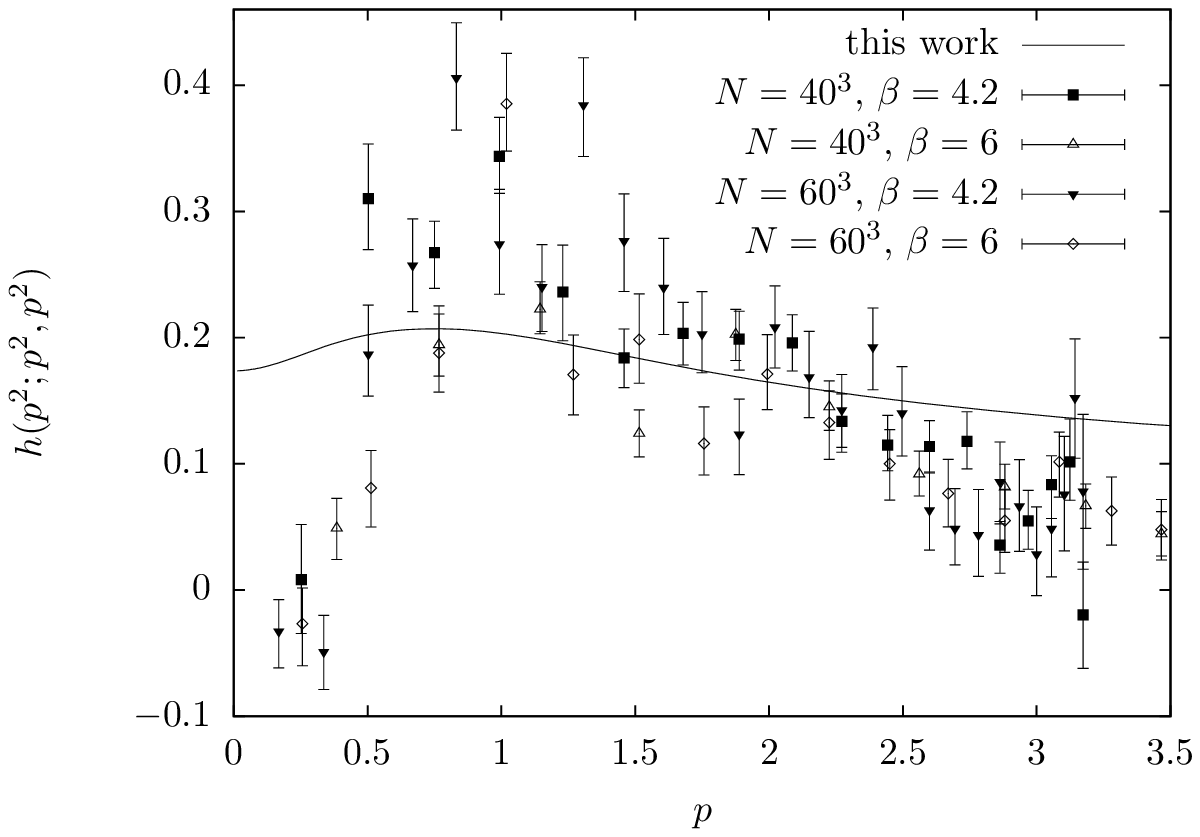}
\caption{\label{fig:lat}Comparison of our result for the form factor of the ghost-gluon
vertex to lattice data from Ref.~\cite{CucMaaMen08} at the symmetric point, $h(p^2;p^2,p^2)$.}
\end{figure}

As explained in the footnote of the introduction, the 3-dimensional Euclidean Yang--Mills
theory in Landau gauge is equivalent to the Hamiltonian approach in Coulomb gauge in $3+1$
dimensions using a specific vacuum wave functional, which is a decent approximation to
the true one in the mid-momentum regime. Therefore we can compare our results with
lattice data from 3-dimensional Landau gauge \cite{CucMaaMen08}, see Fig.~\ref{fig:lat}.
The vertex there displays a maximum in the mid-momentum regime which is somewhat stronger
than in our work; furthermore, the form factor from the lattice seems
to approach zero or even a small negative value in the IR.
Qualitatively similar results have been obtained in 3- and 4-dimensional Landau gauge
both in the continuum \citep{Sch+05} and on the lattice \citep{CucMenMih04}.

%%%%%%%%%%%%%%%%%%%%%%%%%%%%%%%%%%%%%%%%%%%%%%%%%%%%%%%%%%%%%%%%%%%%%%%%%%%%%%%%%%%%%%%%%
%%%%%%%%%%%%%%%%%%%%%%%%%%%%%%%%%%%%%%%%%%%%%%%%%%%%%%%%%%%%%%%%%%%%%%%%%%%%%%%%%%%%%%%%%

\section{\label{sec:conc}Conclusions}

We have investigated the ghost-gluon vertex of the Hamiltonian approach to Yang--Mills
theory in Coulomb gauge. The DSE for the ghost-gluon vertex was solved in one-loop
truncation using the non-perturbative ghost and gluon propagators obtained previously in
the variational approach assuming a bare ghost-gluon vertex. The dressing of the ghost-gluon
vertex was found to increase the vertex between 15\% and 25\% in the IR and to vanish
asymptotically in the UV. Since the gluon propagator obtained in the
variational approach with a bare ghost-gluon vertex is qualitatively similar to that
obtained on the lattice \citep{BurQuaRei09} (which contains the full dressing of
vertices) we expect that a fully self-consistent solution of the coupled DSEs for the
propagators and vertices yield similar results for the ghost-gluon vertex as obtained in
the present paper. Finally it will be interesting to extend the present studies to
finite temperatures \citep{ReiCamSzc11} to see whether there is a similar change of the
dressing of the ghost-gluon vertex at high temperatures as observed in the FRG approach
in Landau gauge \citep{FisPawX}.

\begin{acknowledgments}
The authors are grateful to P.\ Watson and M.\ Quandt for useful discussions and for a
critical reading of the manuscript. They also thank A.\ Maas for providing the lattice
data from Ref.~\cite{CucMaaMen08}.This work has been supported by the Deutsche
Forschungsgemeinschaft under contract no.\ DFG-Re856/6-3.
\end{acknowledgments}

%\bibliographystyle{apsrev4-1}
%\bibliographystyle{h-physrev5}
%\bibliography{biblio}

\end{document}